\documentclass[twocolumn,           % Format : preprint, twocolumn
               showpacs,            % Pacs : showpacs, noshowpacs
               preprintnumbers,     % Preprint: preprintnumbers,
               aps,                 % Society: ...
               prd,                 % Journal Style : pra, prb, prc, prd, pre,
               a4paper,             % Size : a4paper, ...
               superscriptaddress,  % Affiliation (Title) : groupedaddress,
               nofootinbib,         % Footnote: footinbib, nofootinbib
               tightenlines,        % Remove additional spaces in a line
               floats,floatfix      % Floating pictures and tables
               ]{revtex4}

\usepackage{graphicx}
\usepackage{subfigure}
\usepackage{latexsym}
\usepackage{amsmath,amssymb}        % amssymb includes amsfonts
\usepackage[draft=false]{hyperref}

\newcommand{\rd}{{\rm d}}

\input colordvi

\begin{document}

%%--- DRAFTCOPY --------------------------------
%% Prints a large "DRAFT" diagonally across each page
%% Does not show up in TeXview
%% \typeout{Prints "DRAFT" on each page; does not show in TeXView}
%% \special{!userdict begin /bop-hook{gsave 200 30 translate
%% 65 rotate /Times-Roman findfont 216 scalefont setfont
%% 0 0 moveto 0.90 setgray (DRAFT) show grestore}def end}
%%------------------------------------------------

%======================================%
%<<<<<<<<<<<< TITLE PAGE >>>>>>>>>>>>>>%
%======================================%

\title{Dynamics of assisted quintessence}
\author{Soo A Kim}
\affiliation{Astronomy Centre, University of Sussex, Brighton BN1 9QH, United 
Kingdom}
\author{Andrew R.~Liddle}
\affiliation{Astronomy Centre, University of Sussex, Brighton BN1 9QH, United 
Kingdom}
\author{Shinji Tsujikawa}
\affiliation{Gunma National College of Technology, 580 Toriba,
Maebashi, Gunma 371 8530, Japan}
\date{\today}
\pacs{98.80.-k \hfill astro-ph/0506076}
\preprint{astro-ph/0506076}

%======================================%
%<<<<<<<<<<<<< ABSTRACT >>>>>>>>>>>>>>>%
%======================================%

\begin{abstract}
We explore the dynamics of assisted quintessence, where more than one
scalar field is present with the same potential. For potentials with
tracking solutions, the fields naturally approach the same values ---
in the context of inflation this leads to the {\em assisted inflation}
phenomenon where several fields can cooperate to drive a period of
inflation though none is able to individually. For exponential
potentials, we study the fixed points and their stability confirming
results already in the literature, and then carry out a numerical
analysis to show how assisted quintessence is realized.  For inverse
power-law potentials, we find by contrast that there is no assisted
behaviour --- indeed those are the unique (monotonic) potentials where
several fields together behave just as a single field in the same
potential. More generally, we provide an algorithm for generating a
single-field potential giving equivalent dynamics to multi-field
assisted quintessence.
\end{abstract}

\maketitle

%======================================%
%<<<<<<<<<<<<<< ARTICLE >>>>>>>>>>>>>>>%
%======================================%

\section{Introduction}

An attractive hypothesis for the fundamental nature of dark energy is
quintessence --- a scalar field evolving in a non-zero potential
energy \cite{quint}. Such modelling has proven highly successful in
implementing inflation models in the early Universe.  It has however
met with much less success in the present Universe, primarily due to
the difficulty of obtaining equations of state close enough to the
cosmological constant value $w = -1$ to satisfy observational bounds
\cite{obsw}. One strategy is to ensure that the quintessence energy
density starts at such a low value that the field only begins evolving
close to the present, but such tuning is hardly more satisfactory than
a cosmological constant. The alternative is for the field to evolve
significantly, ideally exploiting a `tracker' behaviour rendering the
late-time evolution almost independent of initial
conditions. Unfortunately however this is viable only for very
particular potentials: amongst monotonic potentials, exponentials do
not give acceleration and power-laws can give negative enough $w$ only
for exponents well below 2. Among the potentials regarded as giving
satisfactory phenomenology, many in fact feature a minimum tuned to
match the observed value of the cosmological constant.

In this paper we investigate whether this situation might be
alleviated by allowing the quintessence to arise from several fields,
which we assume to have the same potential energies. While our
interest is primarily phenomenological, we note that such situations
may arise from higher-dimensional theories; in fact there are many
dynamical modulus fields in string theory corresponding to the size of
compactified dimensions.  In the context of early Universe inflation,
such a set of fields has been shown to give the phenomenon of {\em
assisted inflation} \cite{LMS}, whereby they may collectively drive
inflation even if each individual field has too steep a potential to
do so on its own, i.e.~yielding an effective equation of state closer
to $w=-1$. It is therefore worth considering the possibility of
assisted quintessence behaviour, in order to see whether it may be
better able to match observations. Assisted quintessence is
particularly attractive in the context of tracking models, as each
field will separately converge onto the tracking solution making it
entirely natural that they all play a dynamical role.

In this paper we will investigate some aspects of assisted
quintessence, mostly restricting ourselves to the simplest case where
each field has the same potential and always assuming there are no
interactions between fields. The most closely-related paper is that of
Blais and Polarski \cite{BP}, who analyzed several multi-field
quintessence models both analytically and numerically, though with a
different focus directed mainly at attempting to realize models where
the acceleration is a transient phenomenon. We will primarily study
the exponential and inverse power-law cases in detail, in the former
case considering different exponents for the two potentials as already
extensively analyzed by Coley and van den Hoogen \cite{CvdH}. In this
paper we will also provide an algorithm for relating assisted
quintessence to an equivalent single-field model under more general
circumstances.

Our scenario is distinct from two types of scenario already
extensively investigated in the literature. One is the double
exponential potential models of Refs.~\cite{BCN,double}, where there
was only a single field (Ref.~\cite{BCN} briefly mentioned a
multiple-field case but not with uncoupled fields). Another is the
two-field models of dark energy which have received some attention
recently as a way of crossing the cosmological constant boundary
($w=-1$) to give rise to a phantom behaviour at late times
\cite{Feng,GPZZ,crossing}. Unlike our case, at least one of those
fields must have a negative kinetic energy.

\section{Assisted quintessence}

If one accepts the possibility of multiple fields, particularly
sharing the same form of potential $V(\phi_i)$, then the idea of
assisted quintessence emerges very naturally provided the potentials
have tracker solutions for at least some values of the fields. Tracker
solutions arise when the fields are initially sub-dominant as compared
to a perfect fluid, the field contribution to the Friedmann equation
then being neglected to give equations of the form
\begin{eqnarray}
H^2 & \simeq &  \frac{8\pi}{3m_{{\rm Pl}}^2} \rho_{{\rm f}}\,, \\
\ddot{\phi}_i & = & -3 H \dot{\phi}_i - \frac{\rd V}{\rd \phi_i} \,,
\end{eqnarray}
where $H$ is the Hubble rate, $m_{\rm Pl}$ is the Planck mass and a
dot denotes the derivative with respect to a cosmic time $t$.  Here
the fluid density $\rho_{{\rm f}}$ might for instance be broken up
into matter and radiation components $\rho_{{\rm m}}$ and $\rho_{{\rm
r}}$. In this set-up, the scalar fields are completely unaware of each
other's existence (their normal channel of communication being via the
Friedmann equation), and hence separately evolve onto the tracker
solution in response to the fluid. At sufficiently-late times one
would therefore have all the $\phi_i$ equal to each other, and hence
of equal potential energy.

For an individual field $\phi$ with potential $V(\phi)$, whether or not 
there is tracking behaviour can be determined from the value of 
the function
\begin{equation}
\Gamma \equiv \frac{V V''}{V'^2} \,,
\end{equation}
where a prime represents the derivative in terms of $\phi$.  Solutions
converge to a tracker provided that it satisfies $\Gamma > 1-
(2-\gamma)/(4+2\gamma)$ where the equation of state is $p_{{\rm f}}=
(\gamma-1)\rho_{{\rm f}}$, the interesting case however being $\Gamma
>1$ which is required for the field energy density to grow relative to
the fluid allowing eventual domination \cite{SWZ}.

The convergence of different fields to the same tracking solution does
not in itself amount to assisted quintessence, as the fields are not
generating any gravitational effect on the background evolution. The
convergence does however set up the initial conditions for such a
behaviour. What is mainly of interest is what happens once the energy
density of the fields, all evolving together, is no longer
subdominant, and that is the situation addressed in the rest of this
paper. We will therefore be considering the full Friedmann equation
\begin{equation}
H^2 = \frac{8\pi}{3m_{{\rm Pl}}^2} \left( \rho_{{\rm f}} + \sum_i
\rho_{\phi_i} \right) \,,
\end{equation}
where $\rho_{\phi_i} \equiv V_i(\phi_i) + \dot{\phi}_i^2/2$ and we assume 
spatial flatness throughout.

%%%%%%%%%%%%%%%%%%%%%%%%%%%%%%%%%%%%%%%%%%%%
\begin{table*}
\begin{center}
\caption{The properties of the critical points in the presence of a
barotropic fluid with $0<\gamma<2$. There are seven discrete points,
while Case 2 corresponds to a circle of points parametrized by an
angle $\theta$. Case 2 is neutrally stable along the circle, but
always unstable in at least one other direction. The last two columns
show the effective energy density and equation of state of the two
scalar fields combined.}  
\small
\begin{tabular}
{ccccccccc}\hline\hline
\colrule
Case& $x_1$ & $y_1$ & $x_2$ & $y_2$ & Existence & Stability &
$\Omega_{\phi}$ &   
$\gamma 
_{\phi}$ \\\hline
1 & 0 & 0 & 0 & 0 & All $\lambda_1$, $\lambda_2$, $\gamma$ & unstable& 0 & 
Undefined \\
2 &$\cos \theta $ & 0 & $\sin \theta$ & 0 & All & unstable & 1 & 2 \\
%%%%%%%%%%%%%%%%%%%%%%%%%%%%%%%%%%%%%%%%%%%%%%%%%%%%%%%%%%%%%%%%%%%%%%%%%%%
%%%%%%%%%%%%%
3&  $\lambda_1/\sqrt{6}$ 
& $(1-\lambda_1^2/6)^{1/2} $  
& 0 
& 0 
& $\lambda_1^2 <6 $ & unstable & 1 & $\lambda_1^2/3$ \\
%%%%%%%%%%%%%%%%%%%%%%%%%%%%%%%%%%%%%%%%%%%%%%%%%%%%%%%%%%%%%%%%%%%%%%%%%%%
%%%%%%%%%%%%%
4 & $\sqrt{6}\gamma/2\lambda_{1}$ 
& $[3(2-\gamma)\gamma/ 2\lambda_1^2]^{1/2} $  
& 0 
& 0 
& $\lambda_1^2 > 3\gamma $ & unstable & $3\gamma/\lambda_1^2$ & 
$\gamma$ \\
%%%%%%%%%%%%%%%%%%%%%%%%%%%%%%%%%%%%%%%%%%%%%%%%%%%%%%%%%%%%%%%%%%%%%%%%%%%
%%%%%%%%%%%%%
5& 0 
& 0 
& $\lambda_2/\sqrt{6}$ 
& $(1-\lambda_2^2/6)^{1/2} $ 
& $\lambda_2^2<6$& unstable & 1 & $\lambda_2^2/3$ \\
%%%%%%%%%%%%%%%%%%%%%%%%%%%%%%%%%%%%%%%%%%%%%%%%%%%%%%%%%%%%%%%%%%%%%%%%%%%
%%%%%%%%%%%%%
6& 0 
& 0 
& $\sqrt{6}\gamma/2\lambda_{2}$ 
& $[3(2-\gamma)\gamma/ 2\lambda_2^2]^{1/2} $  
& $\lambda_2^2 > 3 \gamma$  & unstable & $3 \gamma/\lambda_2^2$ & 
$\gamma$ \\
%%%%%%%%%%%%%%%%%%%%%%%%%%%%%%%%%%%%%%%%%%%%%%%%%%%%%%%%%%%%%%%%%%%%%%%%%%%
%%%%%%%%%%%%%
7& $\lambda_{\rm eff}^2/\sqrt{6}\lambda_1$ 
& $\frac{\lambda_{\rm eff}}{\lambda_1}(1-\lambda_{\rm 
eff}^2/6)^{1/2}$ 
& $\lambda_{\rm eff}^2/\sqrt{6}\lambda_2$ 
& $\frac{\lambda_{\rm eff}}{\lambda_2} (1-\lambda_{\rm 
eff}^2/6)^{1/2}$ 
& $\lambda_{\rm eff}^2 <6$  
& stable for 
$\lambda_{\rm eff}^2 < 3\gamma  $ & 1 & 
$\lambda_{\rm eff}^2/3$ \\
%%%%%%%%%%%%%%%%%%%%%%%%%%%%%%%%%%%%%%%%%%%%%%%%%%%%%%%%%%%%%%%%%%%%
& & & & & & unstable for $3\gamma < \lambda_{\rm eff}^2  < 6$ &  &  \\
%%%%%%%%%%%%%%%%%%%%%%%%%%%%%%%%%%%%%%%%%%%%%%%%%%%%%%%%%%%%%%%%%%%%%%%%%%%
%%%%%%%%%%%%%%
8& $\sqrt{6}\gamma/2\lambda_{1}$ 
& $[3(2-\gamma)\gamma/ 2\lambda_1^2]^{1/2} $  
& $\sqrt{6}\gamma/2\lambda_{2}$ 
& $[3(2-\gamma)\gamma/ 2\lambda_2^2]^{1/2} $  
&$ \lambda_{\rm eff}^2 > 3\gamma $ & 
stable & $3\gamma/\lambda_{\rm eff}^2 $ & 
$\gamma$ \\
%%%%%%%%%%%%%%%%%%%%%%%%%%%%%%%%%%%%%%%%%%%%%%%%%%%%%%%%%%%%%%%%%%%%%%%%%%%
%%%%%%%%%%%%%%
\hline\hline
\label{table:bw}
\end{tabular}
\end{center}
\end{table*}
%%%%%%%%%%%%%%%%%%%%%%%%%%%%%%%%%%%%%%%%%%%%

\section{Exponential potentials}

We consider two fields $\phi_1$ and $\phi_2$ each with a separate exponential 
potential 
\begin{eqnarray}
V(\phi_1,\phi_2) &=& A e^{-\lambda_1\kappa\phi_1} + B 
e^{-\lambda_2\kappa\phi_2} \\
&\equiv &V_1(\phi_{1})+V_2(\phi_{2})\,, \nonumber
\end{eqnarray}
where $\kappa^2 \equiv 8\pi/m_{{\rm Pl}}^2$. For generality, we will
allow the potentials to have different slopes.

\vspace{0.5cm}

\subsection{Assisted quintessence solutions}

For the case where no matter is present, this system is exactly the
original assisted inflation scenario of Liddle et al.~\cite{LMS},
where the multiple fields evolve to give dynamics matching a
single-field model with
\begin{equation}
\label{lameff}
\frac{1}{\lambda_{{\rm eff}}^2} = 
\frac{1}{\lambda_1^2} + \frac{1}{\lambda_2^2} \,.
\end{equation}
For a single-field potential $V(\phi)=V_{0}e^{-\lambda \kappa \phi}$
the scale factor evolves as $a \propto t^p$, where $p=2/\lambda^2$.
Then the multi-field case Eq.~(\ref{lameff}) corresponds to an
effective power-law index given by $\mbox{$p_{{\rm eff}}\equiv
2/\lambda_{\rm eff}^2=\sum p_i$}$ \cite{LMS}. The expansion rate is
therefore more rapid the more fields there are. A particularly
comprehensive analysis of multiple fields in exponential potentials
has been given by Collinucci et al.~\cite{CNV}.

As it happens, this assisted behaviour continues to be valid in the
presence of a perfect fluid with equation of state $p =
(\gamma-1)\rho$. This was first noted by Coley and van den Hoogen
\cite{CvdH}, who also allowed for the possibility of spatial
curvature. This is because the method used to relate multi-field
dynamics to an equivalent single-field dynamics in Ref.~\cite{LMS} is
a property of the scalar field sector alone, being valid in the
presence of any other matter sources.

\subsection{Critical points and stability}

The critical points for this system were classified by Coley and van
den Hoogen \cite{CvdH}, and we will reiterate their results only
briefly before embarking on some numerical analysis of the evolution
towards them. In Ref.~\cite{CvdH}, spatial curvature was included as a
degree of freedom and the fluid assumed to have $\gamma > 1$, whereas
we assume spatial flatness and $\gamma$ in the wider range $0$ to $2$.

To study the critical point structure and stability of the system, it
is convenient to introduce the following dimensionless quantities
\cite{CLW}
\begin{equation}
x_i \equiv \frac{\kappa \dot{\phi}_i}{\sqrt{6}H}, \; y_i \equiv 
\frac{\kappa \sqrt{V_i}}{\sqrt{3}H}, \quad i=1,2 \,.
\end{equation}
Then we obtain
\begin{widetext}
\begin{eqnarray}
\frac{\rd x_i}{\rd N} &=&  -3x_i +\lambda_i\sqrt{ \frac{3}{2}} \, y_i^2 + 
\frac{3}{2} x_i \left[2 x_1^2 + 
2   x_2^2 + \gamma (1-x_1^2-y_1^2- x_2^2-y_2^2)\right] \qquad i=1,2\,,  \\
\frac{\rd y_i}{\rd N} &=& -\lambda_i \sqrt{\frac{3}{2}}\, x_iy_i +
\frac{3}{2}  
y_i \left[2 x_1^2 + 2 x_2^2 
+ \gamma (1-x_1^2-y_1^2- x_2^2-y_2^2)\right] \qquad \qquad i=1,2\,, \\
\frac{1}{H}\frac{\rd H}{\rd N} &=& -\frac{3}{2}\left[2 x_1^2 + 2 x_2^2
+ \gamma (1-x_1^2-y_1^2- x_2^2-y_2^2)\right]\,,
\end{eqnarray}
\end{widetext}
where $N \equiv \ln a$, together with the constraint
\begin{equation}
\label{constraint}
x_1^2 + y_1^2 + x_2^2 + y_2^2 + \frac{\kappa^2 \rho_{{\rm f}}}{3H^2} =1\,.
\end{equation}
We define the density parameters $\Omega_{\phi_{i}}$ and the equation
of state $w_{\phi_{i}}$ for scalar fields, as
\begin{equation}
\Omega_{\phi_{i}}=x_{i}^2+y_{i}^2\,,~~~
w_{\phi_{i}}=\frac{x_{i}^2-y_{i}^2}{x_{i}^2+y_{i}^2}\,,
~~~~~i=1,2\,.
\end{equation}
We note that the density parameter for matter is given by
$\Omega_{m}\equiv \kappa^2 \rho_{{\rm f}}/3H^2=
1-\Omega_{\phi_{1}}-\Omega_{\phi_{2}}$ from Eq.~(\ref{constraint}).

These equations are in fact valid for any uncoupled potentials, with
$\lambda_i$ defined by
\begin{equation}
\lambda_i \equiv -\frac{1}{\kappa V_i} 
\frac{\rd V_{i}}{\rd \phi_{i}}\,.
\end{equation}
Our case of exponential potentials corresponds to $\lambda_i$ both
constant. It is straightforward to extend our analysis to the case of
a dynamically changing $\lambda$ as studied in single-field models in
Ref.~\cite{MP}.

The classification of critical points and their stability is given in
Table~\ref{table:bw}, and agrees with results in
Ref.~\cite{CvdH}.\footnote{We note that a full analysis has also been
carried out for the case where one of the fields is of phantom type
(opposite sign of the kinetic energy) \cite{GPZZ}, where the late-time
behaviour is domination by the phantom field.}  There are eight types
of critical point, seven being discrete points and one a circular
locus of critical points. They are readily compared to the five types
of critical point found for the single-field system in
Ref.~\cite{CLW}. Our cases 1, 2 with $\theta = 0$ and $\pi$, 3 and 4
correspond to the five points in the single-field case for the field
$\phi_1$, with $\phi_2$ playing no role. Our cases 1, 2 with $\theta =
\pi/2$ and $3\pi/2$, 5 and 6 are the same solutions for the $\phi_2$
field. Finally, cases 7 and 8 are critical points in which both fields
play a role, and which have no direct analogue with the single-field
case.

In the single-field case, the stable late-time attractor is either
scalar-field dominated (case 3 or 5) or a scaling solution (case 4 or
6) depending on the relative values of $\lambda$ and $\gamma$
\cite{CLW}. Once a second field is added, the new degrees of freedom
always render those solutions unstable. The late-time attractors
instead become either the assisted scalar-field dominated solution
case 7 (for $\lambda_{{\rm eff}}^2 < 3 \gamma$) or the assisted
scaling solution case 8 (for $\lambda_{{\rm eff}}^2 > 3 \gamma$).

To compare our results with Ref.~\cite{CvdH}, we note that under their
assumptions the curvature always dominates the fluid at late times,
and behaves like a $\gamma = 2/3$ fluid. Accordingly, they always find
solutions with non-zero fluid density to be unstable to eventual
curvature domination, whereas our assumption of spatial flatness
renders them stable. Otherwise, the classification and stability shown
in our Table matches their Tables I and III.  Additionally they
correctly describe the spatially-flat case in their text.

\subsection{Multi-field phenomenology}

For the exponential potential, the assisted quintessence phenomenon
does indeed exist, with the extra fields being equivalent to a flatter
single-field potential. Unfortunately this result does not seem
particularly useful phenomenologically, as the scaling solutions do
not give acceleration as required by observations, while the
scalar-field dominated solutions would long ago have made the matter
density negligible.

%%%%%%%%%
\begin{figure}
\centering
\includegraphics[width=7.5cm]{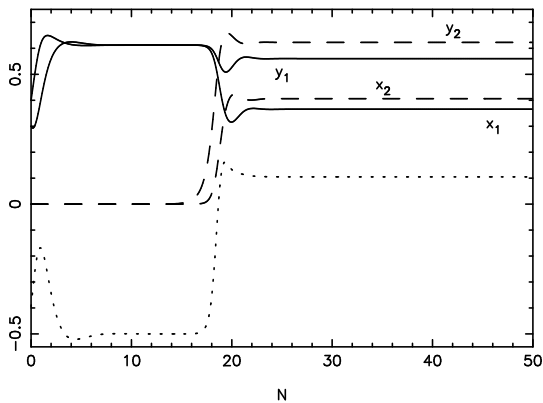}
\caption{A numerical integration for the case $\lambda_1 = 2$,
$\lambda_2 = 1.8$ and $\gamma = 1$, with initial conditions chosen so
that $\phi_1$ dominates.  Field 1 is in solid and field 2 dashed. The
solution initially approaches the single-field scaling solution (case
4), but this becomes unstable once the second field becomes important,
the late-time attractor being the multi-scalar-field dominated
accelerating solution (case 7, $\lambda_{{\rm eff}} = 1.34$). The
dotted line shows $-q=\ddot{a}/aH^2$, where $q$ is the deceleration
parameter.}
\label{fig1}
\end{figure}
%%%%%%%%%% 

There is however one scenario that might be of interest, which is to
imagine there are a large number of exponential potentials with
different initial conditions. As the Universe evolves, more and more
fields would join the assisted quintessence attractor, reducing
$\lambda_{{\rm eff}}$. Eventually, this could switch the attractor
from the scaling regime $\lambda_{{\rm eff}}^2 > 3 \gamma$ into the
regime of late-time scalar field dominance $\lambda_{{\rm eff}}^2 < 3
\gamma$ \cite{CvdH}.

Figure~\ref{fig1} shows a two-field example of such evolution, with
the single-field scaling solution becoming unstable as the second
field becomes important, switching the evolution into late-time scalar
field domination.  {}From Table I the first scaling regime corresponds
to case 4 ($x_1=y_1=\sqrt{3/8}$, $x_2=y_2=0$) with no acceleration,
whereas the final stable attractor is case 7 ($x_1=0.365$,
$y_1=0.560$, $x_2=0.406$, $y_2=0.623$) with acceleration.  We checked
that the values of the fixed points agree very well with numerical
results.

%%%%%%%%%
\begin{figure}
\centering
\includegraphics[width=7.5cm]{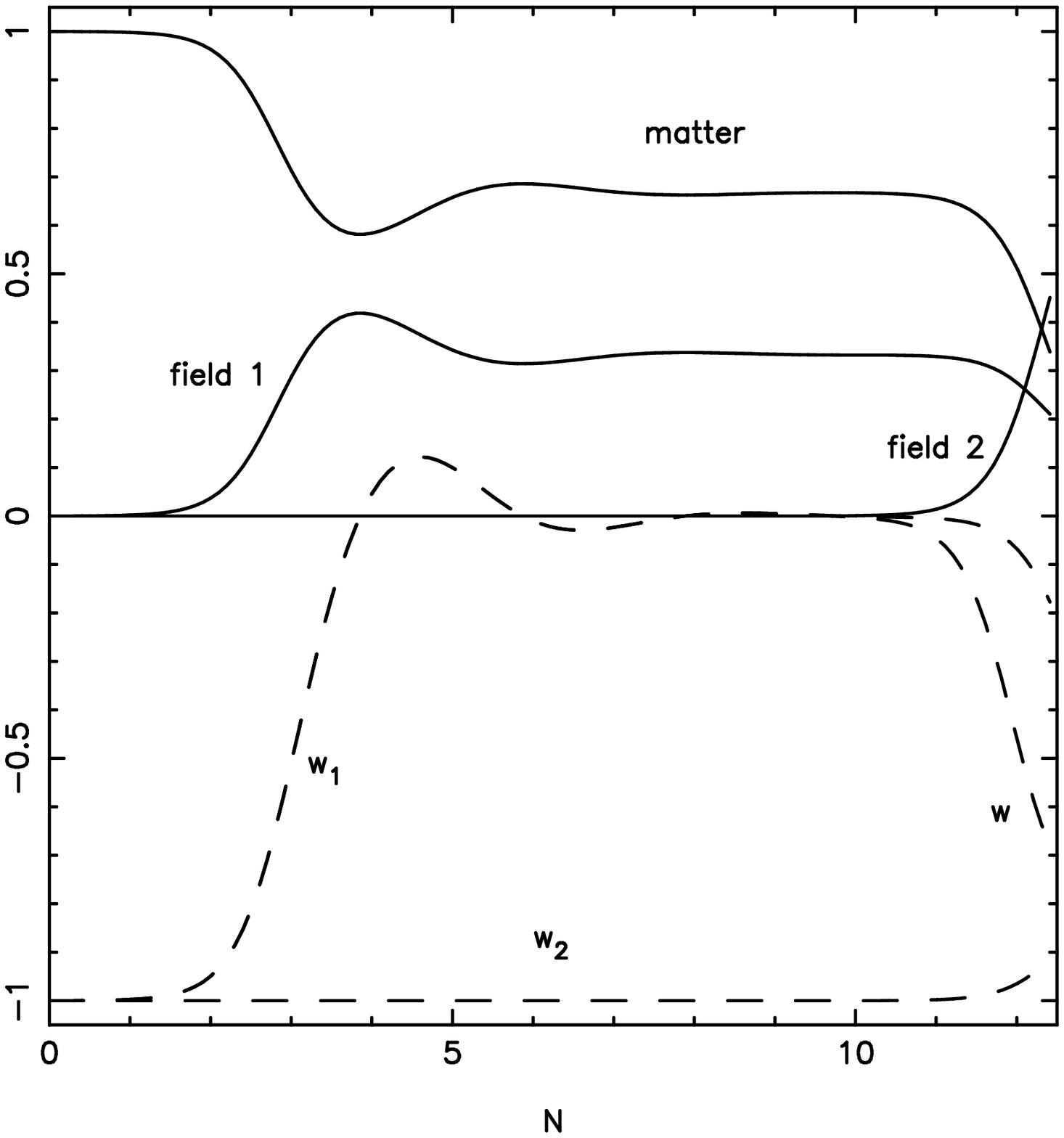}
\includegraphics[width=7.5cm]{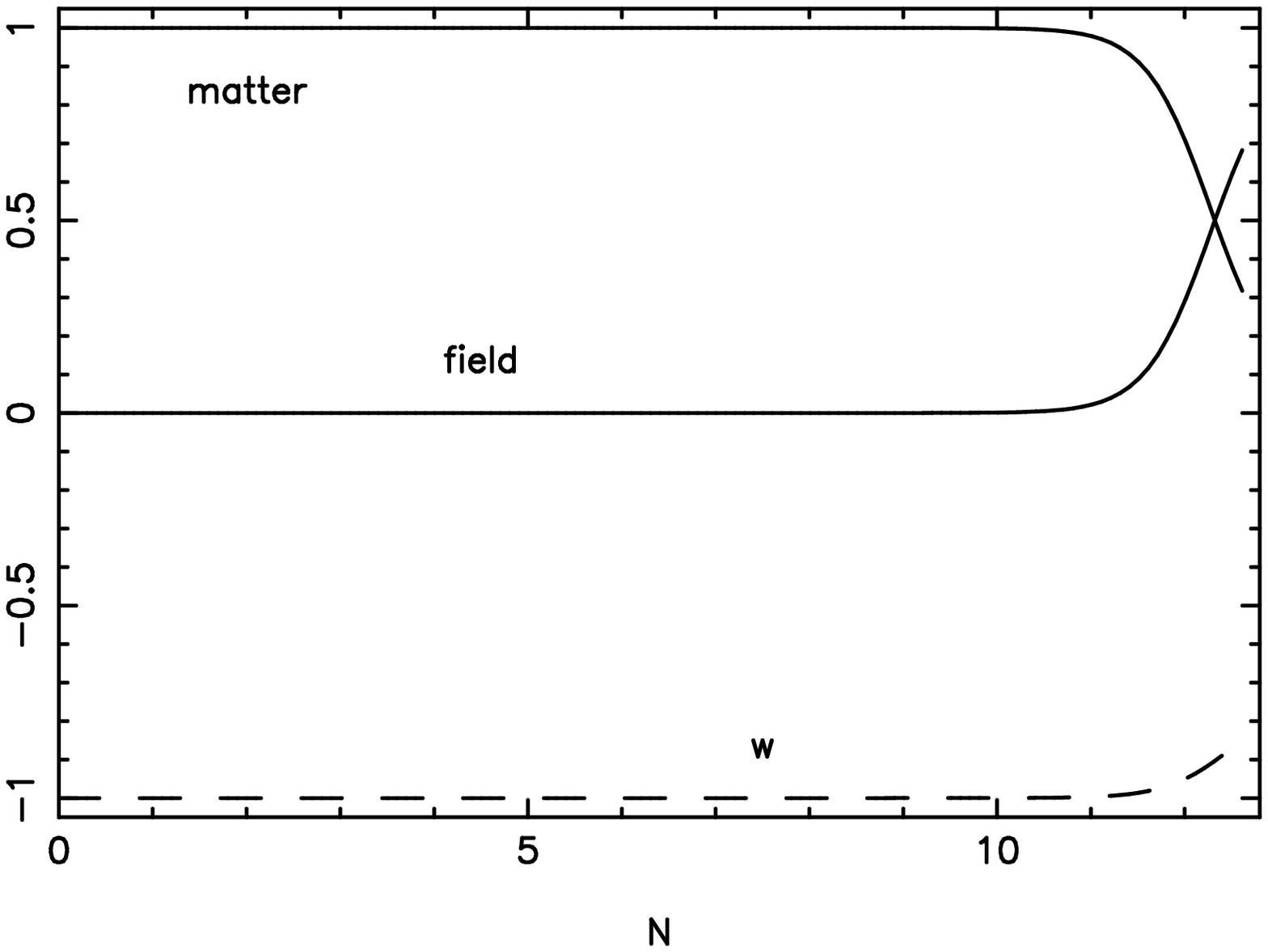}
\caption{A comparison of two-field evolution for $\lambda_1 = 3$ and
$\lambda_2 = 1$ (top) and one-field evolution for $\lambda=1$ (bottom)
.  The density parameters $\Omega_{\phi_{i}}$ and $\Omega_{m}$ are
shown in solid lines and the equation of state parameters
$w_{\phi_{i}}$ and $w$ in dashed lines.  The region around $6<N<11$
corresponds to the scaling regime in which the energy density of the
field $\phi_1$ decreases proportionally to that of the fluid.  The
present epoch is defined in each simulation as the time when the
matter density drops to 0.3 of critical.}
\label{fig2}
\end{figure}
%%%%%%%%%% 

However even if the initial conditions were fine-tuned to bring this
transition into the recent past, it is hard to see how the equation of
state $w$ could reach a sufficiently-negative value to be
observationally viable, current observations indicating roughly $w
\equiv \gamma-1< -0.8$ \cite{obsw}.  In fact, if we compare the
two-field scenarios with a single-field scenario in which an
accelerated expansion occurs at late times ($\lambda<\sqrt{2}$), we
find that typically the two-field models give a larger present value
of $w$ (here $w$ is the equation of state of scalar fields including
the contribution of both $\phi_{1}$ and $\phi_{2}$).
Figure~\ref{fig2} compares a two-field model with $\lambda_1 = 3$ and
$\lambda_2 = 1$ against a single-field model with $\lambda=1$, in both
cases with the field with slope 1 starting with a low value and coming
to dominate only at the present epoch (the `thawing' regime as
according to Ref.~\cite{CL}).  We see that the equation of state of
the thawing field alone is indeed closer to $-1$ at the present
(identified as when the matter density is 0.3 of critical) than in the
single-field case, but the combined $w$ of the two fields is larger.

\section{Inverse power-law potentials}

We now consider the inverse-power law case $V(\phi) = V_0
\phi^{-\beta}$, where we will take the same exponent and normalization
for each field. Here one might hope that the assisted phenomenon would
give rise to an effective $\beta_{{\rm eff}}$ which is smaller than
the individual $\beta$, so that observationally-viable models can be
achieved in steeper potentials than the single-field
case. Unfortunately that turns out not to be true, as we now see.

These potentials are favoured because they exhibit tracker solutions,
and it is therefore legitimate to suppose that after some early time
we can take $\phi_1 = \phi_2 = \ldots \equiv \phi$. The Friedmann and
scalar wave equations become
\begin{eqnarray}
H^2 & = & \frac{8\pi}{3m_{{\rm Pl}}^2} \left( \rho_{{\rm f}} + n V_0 
\phi^{-\beta} + \frac{n}{2} \dot{\phi}^2 \right) \,;\\
\ddot{\phi} & = & - 3 H \dot{\phi} + \beta V_0 \phi^{-\beta-1} \,,
\end{eqnarray}
where $n$ is the number of fields. This is not yet equivalent to a
single-field model, but can be made so (loosely following the method
of Ref.~\cite{LMS}) by the redefinitions
\begin{equation}
\chi = \sqrt{n} \, \phi \quad ; \quad W_0 = \sqrt{n^\beta} \, n V_0 \,,
\end{equation}
which then gives the equations of a single-field model with potential 
$W(\chi)=W_0 \chi^{-\beta}$.

This rescaling indicates that the assisted behaviour renormalizes the
amplitude of the potential in this case, but does {\em not}
renormalize its exponent. In such models the amplitude has to be
adjusted in order to give the present-day value of the matter density
$\Omega_{{\rm m}}$, and having done that the multi-field system then
has identical dynamics to the single-field model, in particular
predicting the same present-day value of the equation of state $w$.

\section{Assisted quintessence dynamics}

We end with a more general construction for analyzing assisted
quintessence, extending the analysis of the previous section to the
case where the potential is arbitrary, but the same for all
fields. Again we assume tracking behaviour so that we can take $\phi_i
= \phi$, giving
\begin{eqnarray}
H^2 & = & \frac{8\pi}{3m_{{\rm Pl}}^2} \left[ \rho_{{\rm f}} + n V(\phi) + 
\frac{n}{2} \dot{\phi}^2 \right] \,,\\
\ddot{\phi} & = & - 3 H \dot{\phi} - \frac{\rd V}{\rd \phi} \,.
\end{eqnarray}
To find an equivalent single-field system, we first note that the
kinetic term in the Friedmann equation forces the correspondence
\begin{equation}
\chi = \sqrt{n} \, \phi \,.
\end{equation}
The equations can then be transformed into single-field form with potential 
\begin{equation}
\label{e:alg}
W(\chi) = n V(\chi/\sqrt{n}) \,,
\end{equation} 
which obviously has the desired effect in the Friedmann equation, but
which also renders the fluid equation into single-field form. This
formula therefore represents an algorithm for finding a single-field
potential $W$ which will generate the same evolution as multiple
fields evolving together in the potential $V$.

For simplicity, we henceforth consider the two-field case, though the 
generalization is straightforward. The correspondence then is
\begin{equation}
\chi = \sqrt{2} \, \phi \quad ; \quad W(\chi) = 2 V(\chi/\sqrt{2}) \,.
\end{equation}
This allows us to ask what condition would have to be satisfied in
order to have no assisted behaviour, i.e.~for $W$ and $V$ have the
same functional form apart from an overall constant. This happens for
potentials obeying the condition
\begin{equation}
V(\chi) = 2 C V(\chi/\sqrt{2}) \,,
\end{equation}
for all $\chi$, where $C$ is a constant. The general solution to this
equation is
\begin{equation}
V(\chi) \propto \chi^{-\beta} \times f(\chi) \,,
\end{equation}
where $f(\chi)$ is any function periodic in $\ln \chi$ with period
$\ln \sqrt{2}$ (i.e.~a Fourier series with this periodicity). While
this can be any of an infinite class of potentials, having such a
periodicity is clearly artificial. The only interesting case therefore
is $f(\chi)$ equals a constant, giving the power-law potential. This
proves that the inverse power-law potentials are the unique monotonic
potentials which do {\em not} exhibit assisted quintessence
behaviour. In all other cases, the equivalent single-field potential
has a different functional form. In the exponential case this
algorithm correctly reproduces Eq.~(\ref{lameff}) for the case
$\lambda_1 = \lambda_2$.

As a final point, we note that under the correspondence
Eq.~(\ref{e:alg}), the tracking parameter $\Gamma_W \equiv WW''/W'^2$
is equal to $\Gamma_V= VV''/V'^2$ (the primes here being derivatives
wrt the arguments $\chi$ and $\phi$ respectively). As one would
expect, the tracking conditions on the multi-field model and its
single-field dynamical equivalent are the same.

\vspace*{0.5cm}
\section{Conclusions}

We have studied various aspects of assisted quintessence
dynamics. Such dynamics arises naturally if there are several fields
with the same potential, provided the potential exhibits tracking
behaviour for at least some stage of its early evolution.

Our most powerful result is Eq.~(\ref{e:alg}), which provides a
general algorithm for finding a single-field model which mimics the
dynamics of a multi-field assisted quintessence model. Applied to
inverse power-law models, it shows that they are the unique
(monotonic) potentials for which there is no assisted behaviour, the
collection of fields behaving as a single field in the same potential
(up to overall normalization). All other potentials will exhibit
assisted behaviour, the exponential potential being an explicit
example.

It should be possible to extend our analysis to the case of more
general dark energy models in which the Lagrangian includes
non-canonical kinematic terms, such as the tachyon, k-essence and
ghost condensate.  For theories whose Lagrangian $p$ is a function of
the field $\phi$ and $X \equiv (\nabla \phi)^2/2$, the existence of
scaling solutions restricts the form of Lagrangian to be
$p=Xg(Xe^{\lambda \phi})$, where $g$ is an arbitrary function and
$\lambda$ is a constant \cite{PT}.  It would be certainly of interest
to investigate whether the assisted behaviour we found for the
canonical scalar field with an exponential potential persists in such
general dark energy models.

It is interesting that the multi-field system can be analyzed so
simply.  Regrettably, however, we have not uncovered any scenarios
where the assisted quintessence phenomenon appears to improve the
situation with regard to the observations. In fact our results show
the contrary; in potentially the most interesting scenario of the
inverse power-law it turns out that there is no assisted quintessence
effect.

%======================================%
%<<<<<<<<<<< ACKNOWLEDGMENTS >>>>>>>>>>%
%======================================%

\begin{acknowledgments}
S.A.K.~was supported by the Korean government, A.R.L.~by PPARC, and
S.T.~by JSPS (No.\,30318802).
\end{acknowledgments}

%======================================%
%<<<<<<<<<<<< BIBLIOGRAPHY >>>>>>>>>>>>%
%======================================%

%%%%%%%%%%%%%%%%%%%%%%%%%%%%%%%%%%%%%%%%%%%%%%%%%%%%%%%%%%%%%%%%%%%%%%%%

\begin{thebibliography}{}
\bibitem{quint} C. Wetterich, Nucl. Phys. {\bf B302}, 668 (1988); B. Ratra 
	and P. J. E. Peebles, Phys. Rev. D{\bf 37}, 3406
	(1988); E. J. Copeland, A. R. Liddle, and D. Wands, Ann. N. Y. Acad.
	Sci. {\bf 688}, 647 (1993); P. G. Ferreira and M. Joyce, Phys. Rev. 
	Lett. {\bf 79}, 4740 (1997), {\tt astro-ph/9707286}, Phys. Rev D 
	{\bf 58}, 023503 (1998), {\tt astro-ph/9711102}; 
	I. Zlatev, L. Wang, and P. J. Steinhardt, 
	Phys. Rev. Lett. {\bf 82}, 896 (1999), {\tt astro-ph/9807002}; 
	A. R. Liddle and R. J. Scherrer, Phys. Rev. D{\bf 59}, 023509 (1999), 
	{\tt astro-ph/9809272}; L. Amendola, Phys. Rev. D{\bf 62},
	043511 (2000), {\tt astro-ph/9908023}; 
	V. Sahni and A. Starobinsky, Int. J. Mod. Phys. \textbf{D9},
	373 (2000), \texttt{astro-ph/9904398};
	T. Padmanabhan, Phys. Rept.  {\bf 380}, 235 (2003), {\tt 
	hep-th/0212290}.
\bibitem{obsw} M. Tegmark et al., Phys. Rev. D{\bf 69}, 103501 (2004), {\tt 
	astro-ph/0310723}; 
	U. Alam, V. Sahni, T. D. Saini, and A. A. Starobinsky,
	Mon. Not. Roy. Astron. Soc. {\bf 354}, 275 (2004), 
	{\tt astro-ph/0311364}; T. R. Choudhury and  T. Padmanabhan, 
	Astron. Astrophys. {\bf 429}, 807 (2005), {\tt
	astro-ph/0311622}, H. K. Jassal, J. S. Bagla, and
	T. Padmanabhan, Mon. Not. Roy. Astron.  
	Soc. {\bf 356}, L11 (2005), {\tt astro-ph/0404378};
	P. S. Corasaniti, M. Kunz, D. Parkinson, E. J. Copeland,
	and B. A. Bassett,
	Phys. Rev. D{\bf 70}, 083006 (2004), {\tt astro-ph/0406608};
	U. Seljak et al., {\tt astro-ph/0407372}.
\bibitem{LMS} A. R. Liddle, A. Mazumdar, and F. E. Schunck, Phys. Rev. D{\bf 
	58}, 061301(R) (1998), {\tt astro-ph/9804177}.
\bibitem{BP} D. Blais and D. Polarski, Phys. Rev. D{\bf 70}, 084008 (2004), 
	{\tt astro-ph/0404043}.
\bibitem{CvdH} A. A. Coley and R. J. van den Hoogen, Phys. Rev. D{\bf 62}, 
	023517 (2000), {\tt gr-qc/9911075}.
\bibitem{BCN} T. Barreiro, E. J. Copeland, and N. J. Nunes,
	Phys. Rev. D{\bf 61}, 127301 (2000), {\tt astro-ph/9910214}.
\bibitem{double} A. A. Sen and S. Sethi, Phys. Lett. B{\bf 532}, 159 (2002), 
	{\tt gr-qc/0111082}; I. P. Neupane,  Class. Quant. Grav. {\bf
	21}, 4383 (2004), {\tt hep-th/0311071};  L. Jarv, T. Mohaupt,
	and F. Saueressig, JCAP {\bf 0408}, 016 (2004), {\tt  
	hep-th/0403063}.
\bibitem{Feng} B. Feng, X. Wang, and X. Zhang, Phys. Lett. B{\bf 607}, 35 
	(2005), {\tt astro-ph/0404224}.
\bibitem{GPZZ} Z.-K. Guo, Y.-S. Piao, X. Zhang, and Y.-Z. Zhang, Phys. Lett. 
	B{\bf 608}, 177 (2005), {\tt astro-ph/0410654}.
\bibitem{crossing} W. Hu, Phys. Rev. D{\bf 71}, 047301 (2005), {\tt 
astro-ph/0410680};
	R. R. Caldwell and M. Doran, {\tt astro-ph/0501104};
	S. Nojiri, S. D. Odintsov, and S. Tsujikawa,
	Phys. Rev. D{\bf 71}, 063004 (2005), {\tt hep-th/0501025};
	X. F. Zhang, H. Li, Y. S. Piao, and X. M. Zhang,
	{\tt astro-ph/0501652}.
\bibitem{SWZ} P. J. Steinhardt, L. Wang, and I. Zlatev, Phys. Rev. D{\bf 59},
	123504 (1999), {\tt astro-ph/9812313}.
\bibitem{CNV} A. Collinucci, M. Nielsen, and T. Van Riet, Class. Quant. Grav. 
	 {\bf 22}, 1269 (2005) {\tt hep-th/0407047}. 
\bibitem{CLW} E. J. Copeland, A. R. Liddle, and D. Wands,
	Phys. Rev. D{\bf 57}, 4686 (1998), {\tt gr-qc/9711068}.
\bibitem{MP}  A. de la Macorra and G. Piccinelli,
	Phys. Rev. D{\bf 61}, 123503 (2000), {\tt hep-ph/9909459};
	S. C. C. Ng, N. J. Nunes, and F. Rosati,
	Phys. Rev. D{\bf 64}, 083510 (2001), {\tt astro-ph/0107321};
	E. J. Copeland, M. R. Garousi, M. Sami, and S. Tsujikawa,
	Phys. Rev. D{\bf 71}, 043003 (2005), {\tt hep-th/0411192}.
\bibitem{CL} R. R. Caldwell and E. V. Linder, {\tt astro-ph/0505494}.
\bibitem{PT} F. Piazza and S. Tsujikawa, JCAP {\bf 0407}, 004 (2004),
	{\tt hep-th/0405054}; S. Tsujikawa and M. Sami,
	Phys. Lett. B{\bf 603}, 113 (2004), {\tt hep-th/0409212};
	B. Gumjudpai, T. Naskar, M. Sami, and S. Tsujikawa,
	JCAP {\bf 0506}, 007 (2005), {\tt hep-th/0502191}.
\end{thebibliography}
\end{document}